\documentclass[preprint,12pt]{elsarticle}
\usepackage{amssymb}

\journal{Physical Review B}

\input{tcilatex}

\begin{document}

\begin{frontmatter}



\title{Nonlinear optical properties in a nanoring: Quantum size and Aharonov-Bohm effect}


\author{Shijun Liang$^{a}$\footnote{E-mail: shijun\_liang@163.com}, Wenfang Xie$^{a }$, H. A. Sarkisyan$^{b,c}$, A. V. Meliksetyan$^{d}$,Huaya Shen$^{e}$}

\address{\small $^{a }$Department of
Physics, College of Physics and Electronic Engineering, Guangzhou
University,
Guangzhou 510006, P.R. China}
\address{\small $^{b }$Russian-Armenian (Slavonic) State University, Yerevan, Armenia}
\address{\small $^{c }$Yerevan State University, Yerevan, Armenia}
\address{\small$^{d}$North Carolina Central University, North Carolina, USA}
\address{\small$^{e}$China Institute for Radiation Protection, Taiyuan 030006, P.R. China}
\begin{abstract}
We have performed calculations of nonlinear optical absorption and nonlinear optical rectification of an exciton in a nanoring in the presence of the magnetic flux. Our results show that one can control properties of nonlinear optical absorption and nonlinear optical rectification of a nanoring by tuning outer and inner radius. In addition, we also find that nonlinear optical properties of a nanoring can be modulated by the magnetic flux through nanoring.
\end{abstract}


\end{frontmatter}



\section{\textbf{Introduction}}
As we know, a charged particle walking through a region where only magnetic flux exists acquires a phase. We call this Aharonov-Bohm effect (ABE). Since Aharonov and Bohm [1] proposed an experiment to assess the manifestations of the electromagnetic potential in the quantum mechanics in 1959, some work [2-4] has been investigated by researchers, and they concluded that all observable phenomena are modulated by the magnetic flux through region, and that these phenomena with respect to $\Phi/\Phi_{0}$ demonstrate period with magnetic flux quantum $\Phi_{0}=hc/e$. With the advance of lithography and semiconductor growth techniques, the nanoring-like devices [5-8] have been realized experimentally. Due to unique character of the nanoring, the self-assembled quantum ring has become the best ideal candidate to observe Aharonov-Bohm (A-B) oscillation [9-11].

    In 1995, Chaplik [12] first put forward predication that the neutral exciton can exhibit A-B oscillator behavior in one-dimensional quantum ring. And subsequently similar results were found in the literature [13] where a short-range interaction potential has been employed. Over the last decade, some researchers had paid their attention to A-B effect of neutral exciton in two-dimensional quantum ring. For instance, Hu $et$ $al$ [9] theoretically studied the magnetic field effects on exciton in an $InAs$ nanoring, They found that the A-B effect of exciton only exists in a finite (but small) width nanoring in the presence of Coulomb correlation. Ribeiro $et$ $al$ [14] experimentally presented a magneto-photoluminescence study of type-II $InP/GaAs$ self-assembled quantum dot, which revealed that A-B type oscillations for neutral excitons when angular momentum of hole ground state ranges from $l_{h}=0$ to $l_{h}=1$, $2$, and $3$. Dias da Silva $et$ $al$ [15] reported impurity effects on the A-B optical signatures of neutral quantum ring magnetoexcitons. Their results, on the one hand, show that energy and oscillator strength of neutral excitons are strongly modulated by the magnetic field. On the other hand, scattering impurity enhances the photoluminescence (PL) intensity on otherwise "dark" magnetic field windows and nonzero PL emission appears for a wide magnetic field range even at zero temperature. Recently Ding $et$ $al$ [10] has investigated the magneto-PL study of a single self-assembled semiconductor quantum ring which is realized by $in$ $situ$ $AsBr_{3}$ etching [16] and experimentally observed the A-B type oscillations in the photoluminescence energy.
As mentioned above, studies about A-B effect are mainly focused on energy and oscillator strength of an exciton in a nanoring. The research into A-B effect on nonlinear optical properties has been rarely reported so far. Understanding nonlinear optical properties of an exciton in a nanoring is of significance for designing optical devices. Hence, it is of interest to study the nonlinear optical properties of an exciton in a nanoring. In the present work, we will focus on studying nonlinear optical absorption and nonlinear optical rectification of an exciton in a nanoring.

The paper is
organized as follows: in section 2 we describe the model and theoretical
framework, section 3 is dedicated to the results and discussions, and
finally, our conclusions are given in section 4.

\section{\textbf{Model and calculations}}
\subsection{\textbf{Exciton in a nanoring}}

Here we are concerned with a nanoring with Winternitz-Smorodinsky confinement potential. In the presence of a magnetic field applied along z-direction, Hamiltonian of single article reads

\begin{equation}
\hat{H}=\frac{1}{2\mu}\left[ \textbf{p+}\frac{e}{c}\textbf{A}\right]
^{2}+V(\textbf{r}),
\end{equation}
In Eq.(1), $\mu$ is effective mass, $e$ is the electron
charge, $c$ is the speed of light in vacuum, $\textbf{A}$ is magnetic vector potential. $V(\textbf{r})$ is given by reference [17]
\begin{equation}
V(\mathbf{r})=\frac{\alpha}{r^{2}}+\beta r^{2}-2\sqrt{\alpha\beta}, \ \ \ \mbox{if $\alpha(\varphi) = constant$}.
\end{equation}

It should be noted that $\alpha$ and $\beta$ are the confinement potential parameters which are connected with the outer radius $R_{2}$ and inner radius $R_{1}$ by the following relations [17]
\begin{equation}
  \beta {R^{2}_{1}} + \frac{\alpha}{R^{2}_{1}}-2\sqrt{\alpha\beta} = U,
  \end{equation}
  \begin{equation}
\beta R^{2}_{2}+\frac{\alpha}{R^{2}_{2}}-2\sqrt{\alpha\beta} = U.
\end{equation}
Where for electron $U = 0.19952meV$ and for hole $U = 0.29928meV$. When the magnetic vector potential $\textbf{A}$ is chosen as $\textbf{A}=\frac{1}{2}Br\hat{\varphi}+\frac{\Phi}{2\pi r}\hat{\varphi}$ and applying the polar coordinates, the Hamiltonian can be rewritten as
\begin{equation}
\hat{H} = -\frac{\hbar^{2}}{2\mu}[\frac{1}{r}\frac{\partial}{\partial r}(r\frac{\partial}{\partial r})+\frac{1}{r^{2}}(\frac{\partial}{\partial \varphi}
+i\frac{\Phi}{\Phi_{0}})^{2}]\nonumber \\
+i\frac{eB\hbar}{2\mu c}(\frac{\partial}{\partial \varphi}+i\frac{\Phi}{\Phi_{0}})-\frac{e^{2}B^{2}r^{2}}{8\mu c^{2}}+V(\textbf{r}),
\end{equation}
where $\Phi_{0}=\frac{hc}{e}$ is magnetic flux quantum.
According to Schr$\ddot{o}$dinger equation
\begin{equation}
H\Psi=E\Psi.
\end{equation}
We can obtain wave function and energy spectrum of single as follows:
\begin{equation}
\Psi(r,\varphi) = \frac{1}{\lambda}\sqrt{\frac{\Gamma[n_{r}+M+1]}{2^{M+1}n_{r}!(\Gamma[M+1])^{2}\pi}}(\frac{r}{\lambda})^{M}e^{-\frac{r^{2}}{4\lambda^{2}}}F(-n_{r},M + 1;\frac{r^{2}}{2\lambda^{2}})e^{im\varphi},
\end{equation}
\begin{equation}
E_{n_{r}m} = (n_{r}+\frac{M}{2}+\frac{1}{2})\hbar\omega -\frac{m-\frac{\Phi}{\Phi_{0}}}{2}\hbar\omega_{c}-2\sqrt{\alpha\beta}.
\end{equation}
Where
 \begin{equation}
 M=\sqrt{(m-\frac{\Phi}{\Phi_{0}})^{2}+\frac{2\alpha\mu}{\hbar^{2}}}.
 \end{equation}
In Eq. (7), $F(x,y;z)$ is the confluent hypergeometric function. $n_{r} =0,1,2,\cdots$ is the radial quantum number, $m =0,\pm1,\pm2,\cdots$ is the magnetic quantum number.

With
\begin{equation}
\omega = \sqrt{\omega_{c}^{2}+\frac{8\beta}{\mu}},
\end{equation}
and
\begin{equation}
\lambda = \sqrt{\frac{\hbar}{\mu\omega}},
\end{equation}
denoting the effective cyclotron frequency and effective magnetic length, respectively.

For an exciton in a nanoring, the Hamiltonian can be written as

\begin{equation}
H = \frac{1}{2\mu_{e}}(\textbf{P}+\frac{e\textbf{A}}{c})^{2}+\frac{1}{2\mu_{h}}(\textbf{P}-\frac{e\textbf{A}}{c})^{2}+V(\textbf{r})-\frac{e^{2}}{\varepsilon|r_{e}-r_{h}|}.
\end{equation}
Here we assume that electron and hole are confined by the same potential. And the last term in right-hand of Eq. (12) is assumed to be perturbation. According to perturbation method, We can obtain the correction from Coulomb interaction to energy[17].

\subsection{\textbf{Oscillator strength}}
The oscillator strength [18] is a very important physical quantity in the study of the optical properties which are related to the electronic dipole-allowed transitions. Generally, the oscillator strength $P_{fi}$ is defined as
\begin{equation}
P_{fi} = \frac{4\mu}{\hbar^{2}}E_{fi}|M_{fi}|^{2},
\end{equation}
 where $E_{fi}=E_{f}-E_{i}$ denotes the energy difference between the initial state and the final state, and $M_{fi} = e<\Psi_{i}|r|\Psi_{f}>$ is the electric dipole moment of the transition from the $\Psi_{i}$ state to the $\Psi_{f}$ state.
\subsection{\textbf{Nonlinear optical rectification and nonlinear optical absorption in a nanoring}}
Based on the density matrix approach and the perturbation
expansion method, the second-order nonlinear optical rectification coefficient is given by [19,20]
\begin{equation}
\chi_{0}^{2} = \frac{4\sigma_{s}M_{fi}^{2}\delta_{fi}(E_{fi}^{2}(1+\frac{T_{1}}{T_{2}})+[(\hbar\omega)^{2}+(\frac{\hbar}{T_{2}})^{2}](\frac{T_{1}}{T_{2}}-1))}{\varepsilon_{0}[(E_{fi}-\hbar\omega)^{2}+(\frac{\hbar}{T_{2}})^{2}][(E_{fi}+\hbar\omega)^{2}+(\frac{\hbar}{T_{2}})^{2}]}.
\end{equation}

In the present work, we also performed calculation of the intensity-dependent resonant peaks of the nonlinear optical absorption and nonlinear optical rectification coefficient which are given by reference [21]
\begin{equation}
\chi_{0,max}^{2}=\xi_{0}(\omega=\omega_{fi},I=0)=\frac{2\sigma_{s}M_{fi}^{2}\delta_{fi}T_{1}T_{2}}{\varepsilon_{0}\hbar^{2}},
\end{equation}
and
\begin{equation}
\alpha_{max}= \alpha(\omega=\omega_{fi},I=0)=\frac{\sigma_{s}E_{fi}M_{fi}^{2}T_{2}}{\hbar^{2}\varepsilon_{0}cn_{r}}.
\end{equation}
Where $\varepsilon_{0}$ is the vacuum permittivity, $n_{r}$ is the refractive index of semiconductor. $\sigma_{s}$ denotes the electron density in the nanoring. In addition, $T_{1}$ and $T_{2}$ are the longitudinal and the transverse relaxation times, respectively. In the calculation, we set the relaxation time at $T_{1}=1ps$, $T_{2}=0.2ps$.
\section{\textbf{Results and Discussions}}
Our calculations are performed for $Al_{0.4}Ga_{0.6}As$ nanoring. The parameters [17,22] chosen in this work are the followings:
$\mu_{e} = 0.067\mu_{0}$, $\mu_{h} = 0.082\mu_{0}$ where $\mu_{0}$ is the free electron mass. $\rho = 3\times 10^{24}m^{-3}$, $n_{r}=3.2$, $\varepsilon=12.9$.

In Fig. 1, we plotted transition energy as a function of inner radius $R_{1}$, outer radius $R_{2}$, respectively, for two different magnetic field $B=5T$, $20T$. From this figure, we can find that transition energy has a continuous increase with respect to inner radius for a fixed outer radius, while that transition energy is reduced for the fixed inner radius as outer radius increases. Here we define width of a nanoring by the difference in size between outer radius and inner radius $\Delta r=R_{2}-R_{1}$. It should be noted that transition energy is not almost affected by the variations of inner or outer radius when $\Delta r$ is greater than $15nm$. Also it can be easily seen that transition energy suddenly becomes very large when $\Delta r$ is very narrow (around $1nm$). The physical origins, for these behaviors, can be interpreted in the following way. For wider ring, exciton moving in the ring can be considered to be two-dimensional, while for narrower ring, one can take exciton as one-dimensional. Two-dimensional nanoring evolves into one-dimensional when width of the nanoring continuously decreases. In addition, in fig. 1, the magnetic field has no influence on the transition energy of an exciton.
Fig. 2 shows oscillator strength as a function of outer radius $R_{2}$, inner radius $R_{1}$, respectively, for two different magnetic field $B=5T$, $20T$. We find from this figure that oscillator strength with respect to inner or outer radius has similar behaviors as shown in Fig. 1. The reasons are that there is a greater probability of transition between the initial and the final state when exciton moves in a narrower ring. In Fig. 3, oscillator strength as a function of magnetic flux $\Phi/\Phi_{0}$ with $R_{1}=5nm$ and $R_{2}=25nm$. It is easily seen from this figure that oscillator strength decreases with increasing $\Phi/\Phi_{0}$. But the Aharonov-Bohm oscillations of exciton is not observed in the nanoring. We can provide this phenomenon with the reason. Electron and hole are confined to the same potential in the nanoring and move over same path, leading to zero in electric dipole moment [23]. Magnetic flux through nanoring has an effect on wave function of electron and hole. Hence, the probability of transition between different states in a nanoring is dependent on magnetic flux.

In Fig. 4, nonlinear optical absorption was presented as a function of inner radius $R_{1}$, outer radius $R_{2}$, respectively, for magnetic field $B=5T$. We see that nonlinear optical absorption is not sensitive to the variations of inner or outer radius for the wider ring, while that nonlinear optical absorption instantly become very huge when $\Delta r$ is reduced to be quite narrow and can be taken as one-dimensional. And Fig. 5 demonstrates that nonlinear optical absorption with respect to magnetic flux $\Phi/\Phi_{0}$ with three different widths of a ring. It is clearly seen that effect of magnetic flux on the nonlinear absorption can be tuned by the width of a ring. So we can tune the outer and inner radius to meet special needs in experiment, which is very significant in designing devices.

In Fig. 6, nonlinear optical rectification was plotted as a function of photon energy with magnetic field for three different $\Delta r$. We note from this figure that the magnitude of resonant peak of nonlinear rectification increases as $\Delta r$ increases. Also, the position of resonant peak of nonlinear optical rectification has a blue shift, which is modulated by width of a ring. In addition, we also observe that the magnitudes of resonant peak are far less than results from literature [21] where parabolic confinement potential is employed to confine electron and hole. And this indicates that the Winternitz-Smorodinsky potential is not realistic in crystal.
Fig. 7 shows nonlinear optical rectification as a function of inner radius $R_{1}$, outer radius $R_{2}$, respectively, for magnetic field $B=5T$. When outer radius is set at $31 nm$, nonlinear optical rectification is negative and decreases with inner radius ranging from $0 nm$ to $5 nm$. And then the nonlinear rectification monotonously increases with increasing inner radius. In addition, we also see that the nonlinear optical rectification monotonously decreases with outer radius for a fixed $R_{1}$. Finally, we shows nonlinear optical rectification as a function of magnetic flux $\Phi/\Phi_{0}$ for $R_{1}=5 nm$ and $R_{2}=50 nm$  in Fig. 8. From this figure, we can see that the nonlinear rectification increases with respect to $\Phi/\Phi_{0}$. And no A-B oscillation is observed in this figure.

\section{\textbf{Summary}}
 We have investigated the effects of magnetic flux, size of quantum ring on transition energy, oscillator strength, nonlinear optical absorption and nonlinear optical rectification of exciton in a nanoring with Winternitz-Smorodinsky confinement potential. Our results show that transition energy, oscillator strength, nonlinear optical absorption and nonlinear optical rectification are strongly affected by width of a nanoring, and that one can control these properties of a nanoring by modulating the outer or inner radius to meet special needs in designing electro-optical devices. In addition, it is found that the magnetic flux can tune transition energy, oscillator strength, nonlinear optical absorption and nonlinear optical rectification of a nanoring.

 In conclusion, the properties of an exciton in nanoring are strongly dependent on width of a ring and magnetic flux. Finally we hope our research can stimulate further studies in nonlinear optical properties of a nanoring.
\section{\textbf{Acknowledgement}}
This work is supported by National Natural Science Foundation of China(under
Grant No. 11074055).

\newpage

\section{caption}

Fig. 1. Transition energy as a function of inner radius $R_{1}$, outer radius $R_{2}$, respectively, for two different magnetic field $B=5T$, $20T$.

Fig. 2. Oscillator strength as a function of inner radius $R_{1}$, outer radius $R_{2}$, respectively, for two different magnetic field $B=5T$, $20T$.

Fig. 3. Oscillator strength as a function of magnetic flux $\Phi/\Phi_{0}$ with $R_{1}=5nm$ and $R_{2}=25nm$.

Fig. 4. Nonlinear optical absorption was presented as a function of inner radius $R_{1}$, outer radius $R_{2}$, respectively, for magnetic field $B=5T$.

Fig. 5. Nonlinear optical absorption with respect to magnetic flux $\Phi/\Phi_{0}$ with three different widths of ring.

Fig. 6. Nonlinear optical rectification was plotted as a function of photon energy with magnetic field for three different widths of ring.

Fig. 7. Nonlinear optical rectification as a function of inner radius $R_{1}$, outer radius $R_{2}$, respectively, for magnetic field $B=5T$.

Fig. 8. Nonlinear optical rectification as a function of magnetic flux $\Phi/\Phi_{0}$ for $R_{1}=5nm$ and $R_{2}=50nm$.
\newpage
\begin{figure}[tbp]
\begin{center}
\includegraphics[scale=1.2]{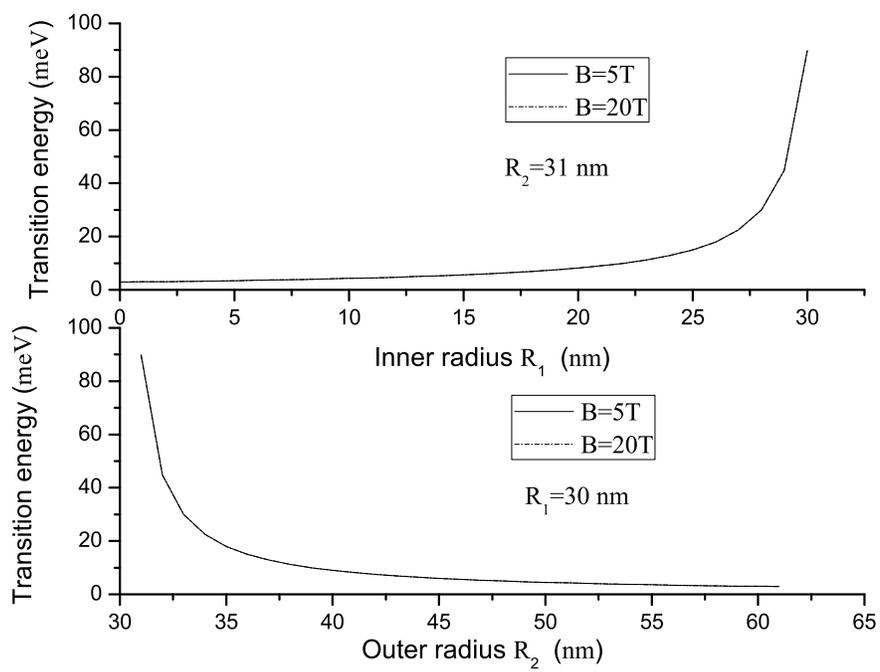}
\end{center}
\caption{Transition energy as a function of inner radius $R_{1}$, outer radius $R_{2}$, respectively, for two different magnetic field $B=5T$, $20T$.}
\end{figure}

\begin{figure}[tbp]
\begin{center}
\includegraphics[scale=1.2]{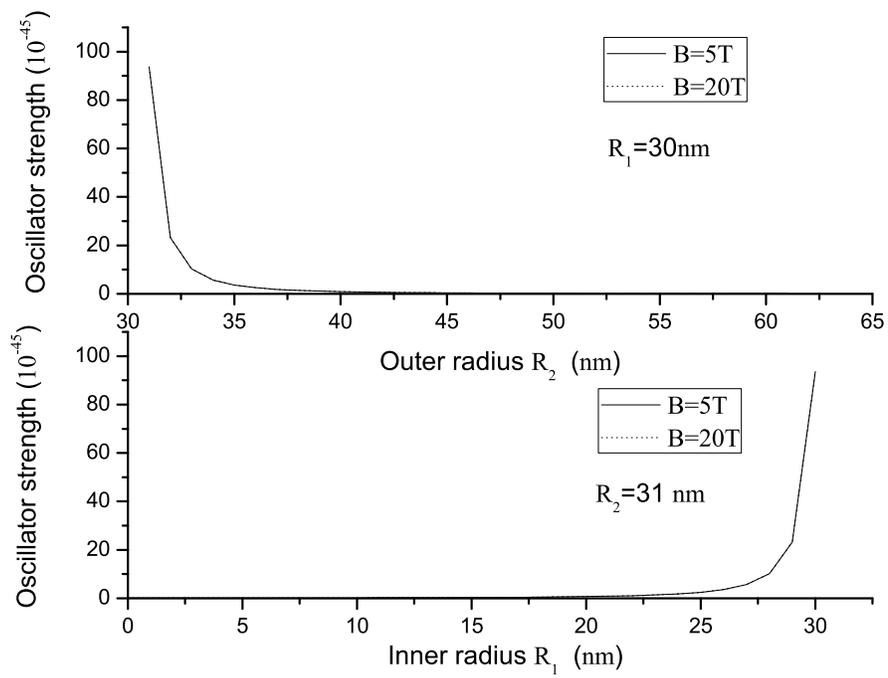}
\end{center}
\caption{Oscillator strength as a function of inner radius $R_{1}$, outer radius $R_{2}$, respectively, for two different magnetic field $B=5T$, $20T$.}
\end{figure}

\begin{figure}[tbp]
\begin{center}
\includegraphics[scale=1.2]{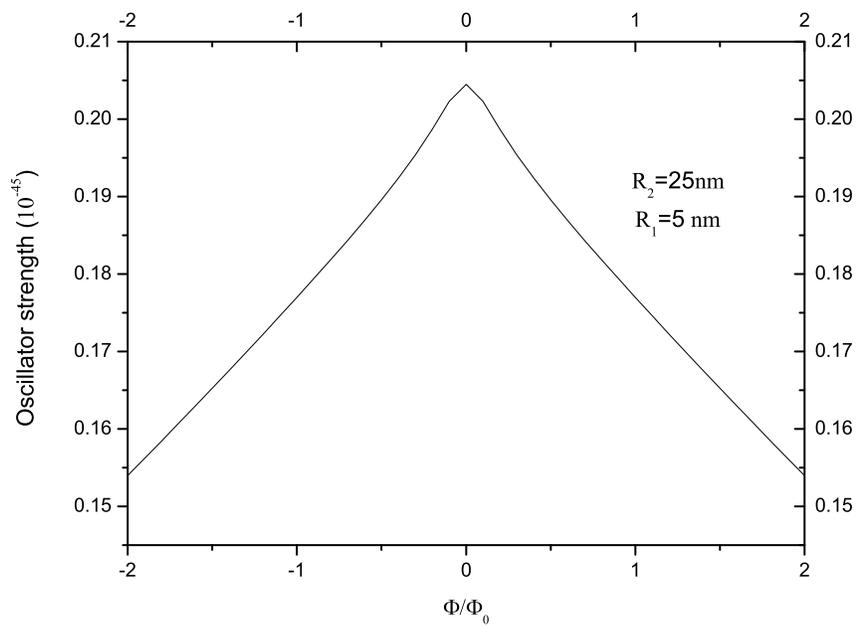}
\end{center}
\caption{Oscillator strength as a function of magnetic flux $\Phi/\Phi_{0}$ with $R_{1}=5nm$ and $R_{2}=25nm$.}
\end{figure}

\begin{figure}[tbp]
\begin{center}
\includegraphics[scale=1.2]{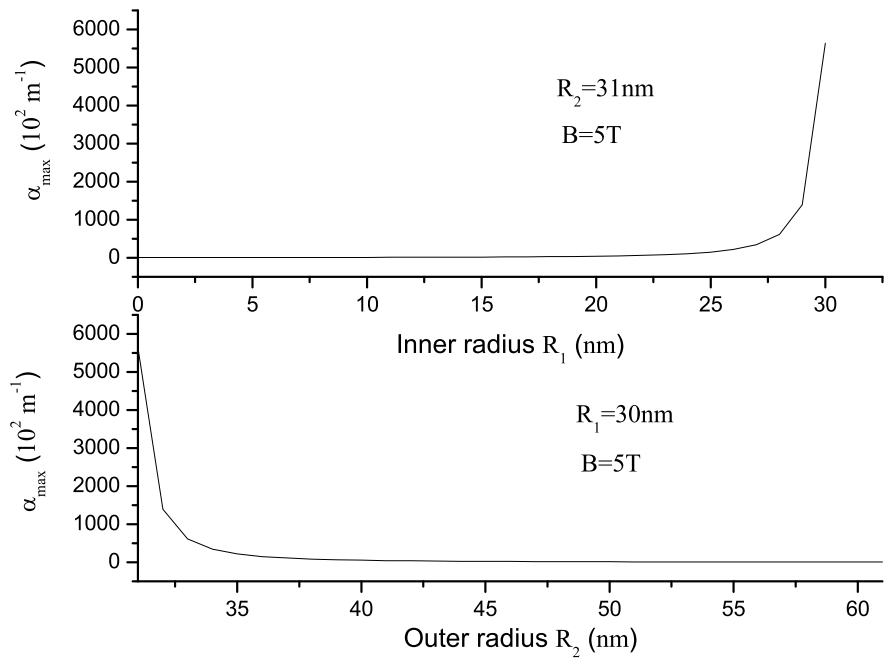}
\end{center}
\caption{Nonlinear optical absorption was presented as a function of inner radius $R_{1}$, outer radius $R_{2}$, respectively, for magnetic field $B=5T$.}
\end{figure}
\begin{figure}[tbp]
\begin{center}
\includegraphics[scale=1.2]{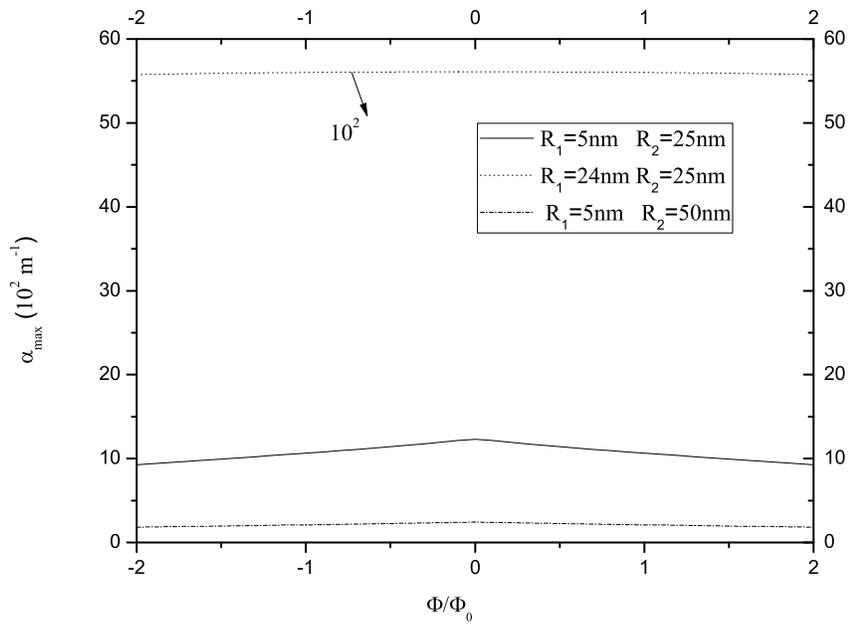}
\end{center}
\caption{Nonlinear optical absorption with respect to magnetic flux $\Phi/\Phi_{0}$ with three different widths of ring.}
\end{figure}
\begin{figure}[tbp]
\begin{center}
\includegraphics[scale=1.2]{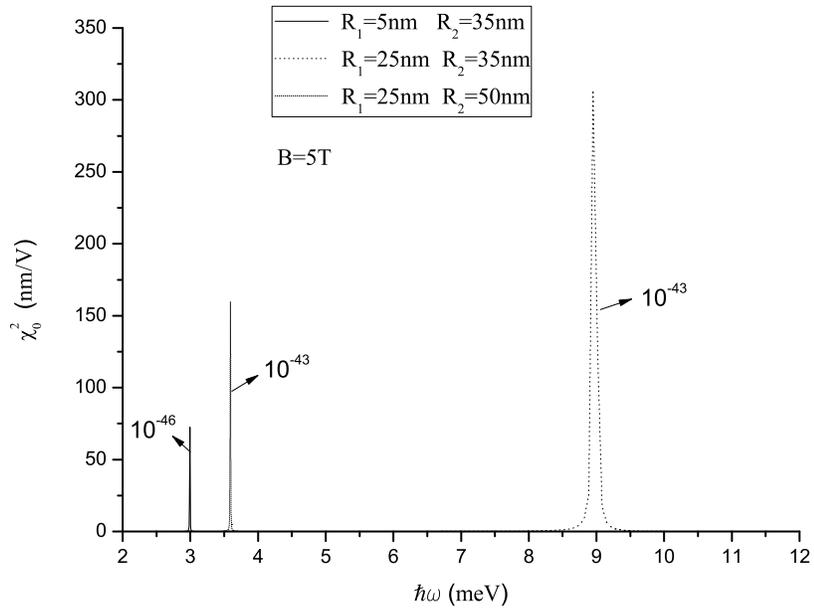}
\end{center}
\caption{Nonlinear optical rectification was plotted as a function of photon energy with magnetic field for three different widths of ring.}
\end{figure}
\begin{figure}[tbp]
\begin{center}
\includegraphics[scale=1.2]{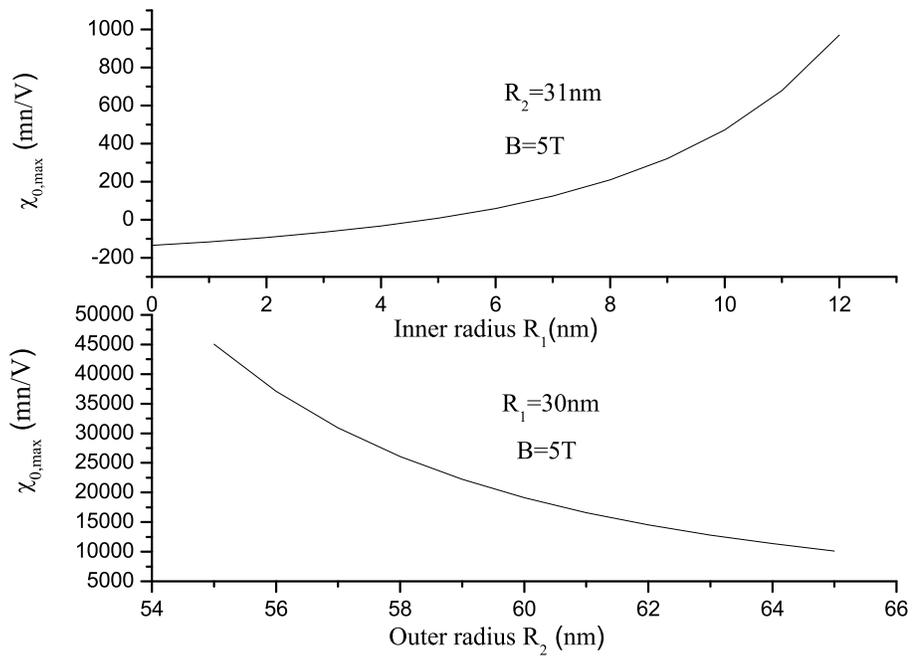}
\end{center}
\caption{Nonlinear optical rectification as a function of inner radius $R_{1}$, outer radius $R_{2}$, respectively, for magnetic field $B=5T$.}
\end{figure}
\begin{figure}[tbp]
\begin{center}
\includegraphics[scale=1.2]{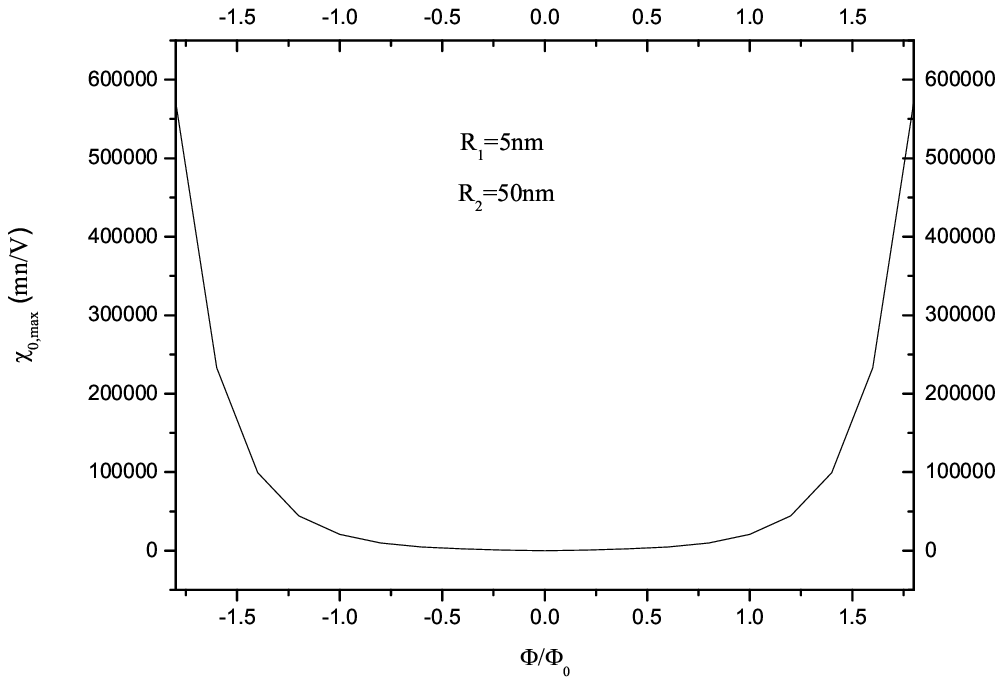}
\end{center}
\caption{Nonlinear optical rectification as a function of magnetic flux $\Phi/\Phi_{0}$ for $R_{1}=5nm$ and $R_{2}=50nm$.}
\end{figure}

\vskip0.5cm \newpage
\begin{thebibliography}{39}
\bibitem{ref1} Y.Aharonov and D.Bohm, Phys. Rev. 115, 485 (1959).
\bibitem{ref2} N. Byers and C.N. Yang, Phys. Rev. Lett. 7, 46 (1961).
\bibitem{ref3}F. Bloch, Phys. Rev. Lett. 21, 1241 (1968).
\bibitem{ref4} M. B$\ddot{u}$ttiker, Y. Imry and R. Landauer, Phys. Lett. 96A, 365 (1983).
\bibitem{ref5} G. Timp, A.M. Chang, J.E. Cunningham, T.Y. Chang, P. Mankiewich, R. Behringer and R.E. Howard, Phys. Rev. Lett. 58, 2814 (1987).
\bibitem{ref6} C.J.B. Ford, T.J. Thornton, R. Newbury, M. Pepper, H. Ahmed, C.T. Foxon, J.J. Harris and C. Roberts, J. Phys. C 21, L325 (1988).
\bibitem{ref7} A. Fuhrer, S. L$\ddot{u}$sher, T. Ihn, T. Heinzel, K. Ensslin, W. Wegsheider and M. Bichler, Nature (London) 413, 822 (2001); Microelectron. Eng. 63, 47 (2002).

\bibitem{ref8} J.B. Yau, E.P.De Poortere and M. Shayegan, Phys. Rev. Lett. 88, 146801 (2002).
\bibitem{ref9} Hui Hu, Jia-Lin Zhu, Dai-Jun Li and Jia-Jiong Xiong, Phys. Rev. B 63, 195307 (2001).

\bibitem{ref10} F. Ding, N. Akopian, B. Li, U. Perinetti, A. Govorov, F.M. Peeters, C.C. Bof Bufon, C, Deneke, Y. H. Chen, A.Rastelli, O.G. Schmidt and V.Zwiller, Phys. Rev. B 82, 075309 (2010).

\bibitem{ref11} R.A. R$\ddot{o}$mer and M.E. Raikh, arXiv:cond-mat/9906314v2 (2000).
\bibitem{ref12} A. Chaplik, Pisma Zh. $\acute{E}$ksp, Teor. Fiz. 62, 855 (1995) [JETP Lett. 62, 900 (1995)].

\bibitem{ref13}R.A. R$\ddot{o}$mer and M.E. Raikh, Phys. Rev. B 62, 7045 (2000).
\bibitem{ref14} E. Ribeiro, A.O. Govorov, W. Carvalho, Jr. and G. Medeiros-Ribeiro, Phys. Rev. Lett. 92, 12 (2004).
\bibitem{ref15}L.G.G.V. Dias da Silva, S.E. U${\i}{\i}$oa and A.O. Govorov, Phys. Rev. B 70, 155318 (2004).
\bibitem{ref16} F. Ding, L. Wang, S. Kiravittaya, E. M$\ddot{u}$ller, A. Rastelli and O.G. Schmidt, Appl. Phys. Lett. 90, 173104 (2007).

\bibitem{ref17}A. K. Atayan, E. M. Kazaryan, A. V. Meliksetyan, H. A. Sarkisyan, Journal of Contemporary Physics. 45, 126 (2010).
\bibitem{ref18} W.F. Xie, Phys. E 43, 49 (2010).
\bibitem{ref19}S. Baskoutas, E. Paspalakis, A.F. Terzis, Phys. Rev. B 74, 153306 (2006).

\bibitem{ref20}E. Rosencher, Ph. Bois, Phys. Rev. B 44, 11315 (1991).

\bibitem{ref21}M. Zalu$\dot{z}$ny, J. Appl. Phys. 74, 4716 (1993).

\bibitem{ref22} W. F. Xie, Journal of Luminescence 131, 943 (2011).
\bibitem{ref23}A. O. Govorov, S. E. Ulloa, K. Karrai, and R. J. Warburton, Phys. Rev. B 66, 081309 (2002).


\end{thebibliography}
\end{document}